\documentstyle[twoside,fleqn,jhuwkshp,psfig]{article}

\newcommand{\chandra}{{\it Chandra} }

\newcommand{\rxte}{{\it RXTE} }

\begin{document}

\title{Highly Ionized Absorption in the X-ray Spectrum of Cyg X-1}
 
\author{H. L. Marshall
\address{MIT Center for Space Research, Cambridge, MA 02139, USA.},
N. S. Schulz$^{\rm a}$, T. Fang$^{\rm a}$,
W. Cui \address{Department of Physics, Purdue University
West Lafayette, IN 47907, USA.},
C. R. Canizares$^{\rm a}$, J. M. Miller$^{\rm a}$, W.H.G. Lewin$^{\rm a}$}

\begin{abstract}
Using the Chandra High Energy Transmission Grating Spectrometer (HETGS), we
have found significant absorption features in the X-ray spectrum of Cyg X-1
taken in the continuous clocking mode.
These features include resonance lines of He-like ions of S, Si, and Mg; the
Ly alpha lines of H-like S, Si, Mg, and Ne; and several lower ionization lines
of Fe {\sc XX}, {\sc XXII}, and {\sc XXIV}.
Preliminary analysis shows that
the lines are resolved in many cases, giving line
widths of order 300 km/s and are redshifted by
460 $\pm$ 10 km/s.  These features are interpreted in the context of an
accreting stellar wind model that is ionized by the X-ray source.
In addition, there are clear absorption features due to neutral Mg, Fe,
and O in the interstellar medium.
\end{abstract}

\maketitle

\section{Introduction}

Cyg X-1 has been observed several times with the
High Energy Transmission Grating Spectrometer (HETGS) on the
\chandra\ X-ray Observatory.  The first observation
\cite{schulz01} shows details
of the structures of the ionization edges from neutral
species in the interstellar medium.  In particular,
the O K and Fe L edges showed significant structure that had
been unobservable at low spectral resolution.
The Ne K edge was unresolved but the nearby continuum
showed subtle features that could be attributed to
Fe {\sc XX}.  There were other weak, unresolved features in the
spectrum that supported in the interpretation that we
are observing Cyg X-1 through a low density ionized gas
that is associated with Cyg X-1.

A second observation with the HETGS
that was taken at a different orbital phase and in
a different instrument mode.  We find many redshifted
absorption features from highly ionized species of
S, Si, Mg, Ne, and Fe.  We present a preliminary analysis
of this observation.
% In a later observation
% \cite{miller02} these lines were also observed but
% are not redshifted.

\section{Chandra HETGS Spectra}

Cyg X-1 was observed on 12 January 2000
(MJD = JD - 2400000.5 = 51555.34 - 5.51).
The orbital phase for this observation was 0.84, based on
a recent orbital ephemeris \cite{brocksopp99}.  For the
first observation \cite{schulz01}, the orbital phase
was 0.93, while in a third
observation, the orbital phase was 0.76, nearly
perpendicular to the line between the compact and
companion stars \cite{miller02}.
The total exposure time was 12677.5 s.
The Advanced CCD Imaging Spectrometer (ACIS-S) detector was used
for readout and was operated in the so-called ``continuous clocking''
(CC) mode.  This mode is described in detail in the \chandra\ Proposer's
Observatory Guide (p. 101).  Briefly,
in this mode, the charge is shifted row by row to the frame
storage buffer at 2.84996 ms intervals.  Imaging information along
the readout direction is lost in favor of improved event timing.

\begin{figure*}[t] % fig.1
\vspace{10pt}
{\psfig{file=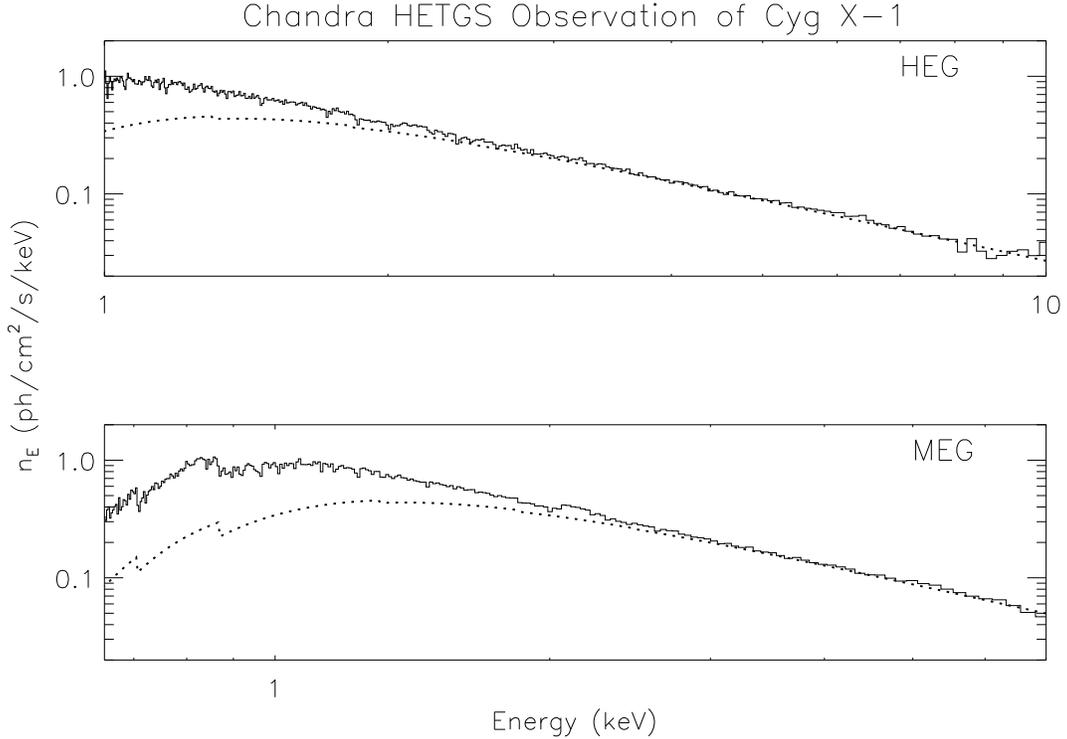,width=6.0in}}
\caption{1-10 keV spectrum of Cyg X-1 from
the \chandra\ HETGS.  Spectra from the HEG (top) and MEG
(bottom) portions of the HETGS are shown separately.
The data have been heavily binned to show the continuum shape.
The model (dotted line) is a simple power law with $\Gamma = 1.73$
absorbed by a neutral column density of $N_H = 6 \times 10^{21}$
cm$^{-2}$ \cite{schulz01}.  A soft excess is attributed to a thermal
component is detected below 2 keV.  Narrow
absorption features are observed in the 1.5-2.5 keV range.
These are more clearly shown in Fig.~\ref{fig:spectrum}.
}\label{fig:hegspec}
\end{figure*}

\begin{figure*}[ht] % fig.2
\vspace{10pt}
{\psfig{file=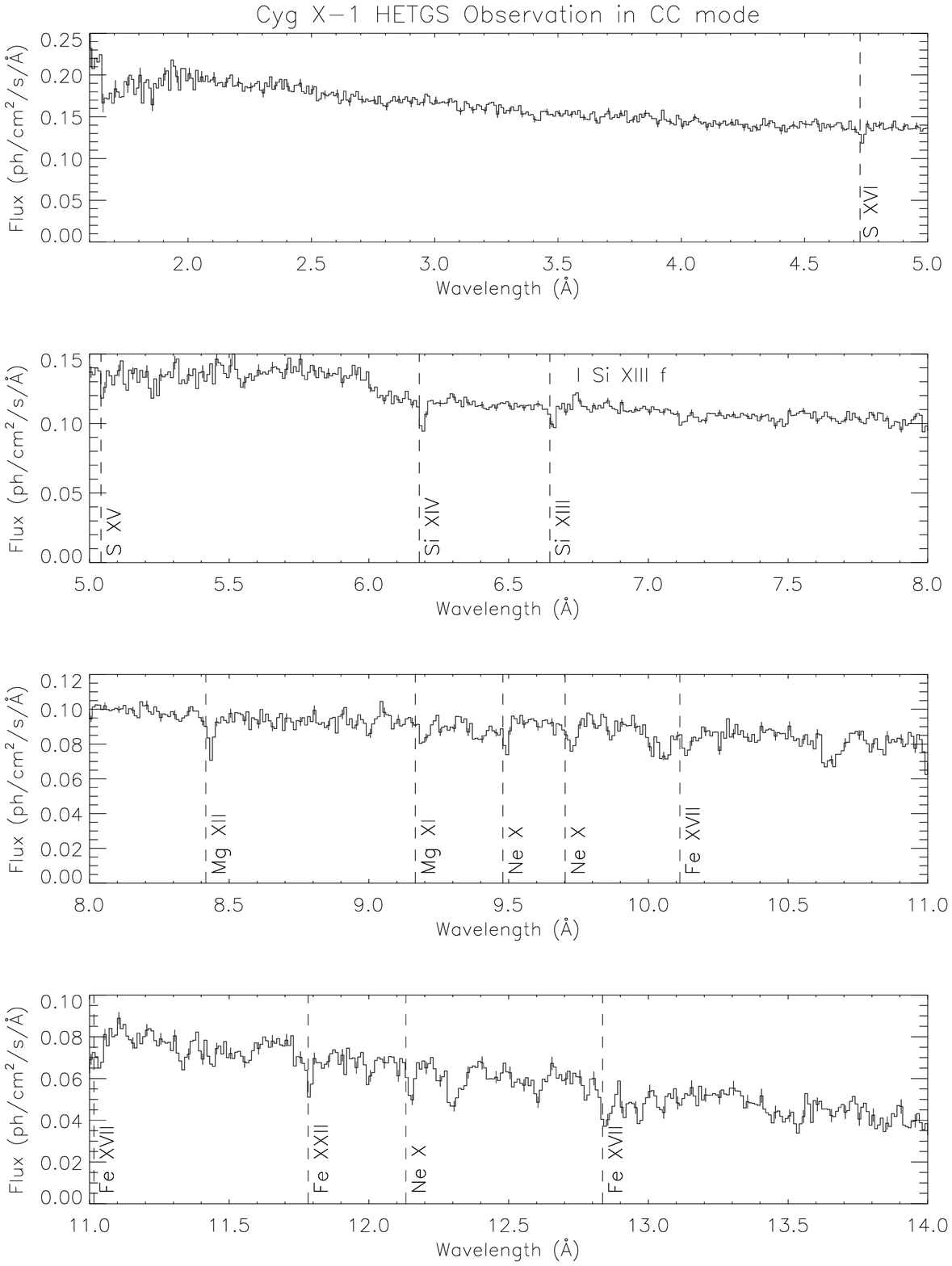,width=6.0in}}
\caption{\chandra\ HETGS spectrum of Cyg X-1.  The MEG
and HEG spectra were combined at a binning of 0.01 \AA.
Note the many absorption lines of highly ionized species,
ranging from Fe {\sc XVII} to S {\sc XVI}.  The vertical
dashed lines mark the rest wavelengths of each line.
The average redshift is 460 $\pm$ 10 km/s.
}\label{fig:spectrum}
\end{figure*}

The \chandra\ HETGS spectrum of Cyg X-1 is
shown in Figs.~\ref{fig:hegspec}
and \ref{fig:spectrum}.  A fit to a simple power law
in the 2-10 keV band gives a photon index of 1.73.
The residuals (Fig.~\ref{fig:hegspec}) show that
there is a soft excess that is usually attributed
to a disk blackbody.  In the first HETGS observation,
$kT$ of the disk was about 200 eV \cite{schulz01}.
In addition to absorption edges from neutral gas
in the interstellar medium, there are
many narrow absorption lines.  The depths of
the Ne K and Fe L absorption edges are consistent with
the edge strengths obtained from the first observation
where $N_H$ was estimated to be
6.2 $\times 10^{21}$ cm$^{-2}$ \cite{schulz01}.

In order to show the absorption lines
more clearly, the HETGS medium energy grating (MEG)
and the high energy grating (HEG) spectra were
combined at a binning of 0.01 \AA
(Fig.~\ref{fig:spectrum}).
There are more than
a dozen narrow absorption lines of highly ionized species,
ranging from Fe {\sc XVII} to S {\sc XVI}.
The lines are redshifted by 300 to 800 km/s
but are generally consistent with the weighted mean:
460 $\pm$ 10 km/s (see Fig.~\ref{fig:lines}). 
The lines are
resolved in HEG spectra (and often in the
lower resolution MEG spectra as well),
giving Doppler line widths of order 300 km s$^{-1}$ (FWHM).
In some cases, blends cause the lines to be broader than
expected.

Practically no Ly $\beta$ lines are
detected while Ly $\alpha$ lines are strong,
so we infer that the lines are not saturated
and that the absorbing gas probably covers more
than $\sim 50$\% of the source.  Assuming complete
coverage, we estimated column densities from fits of Gaussian
models jointly to the original MEG and HEG data.
The results are shown in Fig.~\ref{fig:lines}.  
The column densities are between $10^{16}$ cm$^{-2}$ and
$10^{17}$ cm$^{-2}$, which, using solar abundances, correspond
to equivalent hydrogen column densities of the order of $10^{20-21}$
cm$^{-2}$. There is trend that the column densities of He-like ions
are systematically lower than the corresponding H-like ions, which 
could indicate that the photoionized plasma has a high temperature
or the emission from these ions originate from different emission zones.
We note that the result for all the Fe {\sc XVII}
transitions are consistent with each other.          

\section{Discussion}

We consider a simple working hypothesis in
which the narrow lines result from
absorption by ionized gas as it is transferred from the
companion star to the black hole.  The companion is a
supergiant O star with a with a radiatively
driven wind with a maximum velocity
of $\sim 1500$ km s$^{-1}$ \cite{cak75}.
Cyg X-1 gravitationally focusses and accretes this wind
from the companion \cite{gb86} although the companion
may be very close to filling its Roche lobe \cite{wen99}.
The wind from the companion is expected to be too cool
($T \sim 10^5$ K, as in Vela X-1 \cite{schulz02})
to ionize S and Si to the H- and He-like states, so the
ionization mechanism is most likely photoionization by
the hard X-ray continuum.  This continuum is thought to
result from Comptonization of the accretion disk's thermal
radiation in an extremely hot corona, so the absorbing
gas must have a much larger physical scale.

In general, the wide range of observed ionization states
(Fe {\sc XVII} to S {\sc XVI}) would not be found in gas
of a single ionization parameter, $\xi \equiv L_x / (n r^2)$,
so $n r^2$ probably varies significantly along the line
of sight.  Lines of these strengths
were not observed in the first observation, so the conditions
under which they are detected may depend on the luminosity
state of the illuminating X-ray source or the orbital phase.
The source was $\times$ 2 brighter during the first observation
so it may be that the higher X-ray luminosity
completely ionized the accreting material.
On the other hand, the orbital phase was 0.93, so the wind
may have been mostly obscured by the companion star.

\begin{figure}[ht] % fig.3
\vspace{10pt}
\centerline{\psfig{file=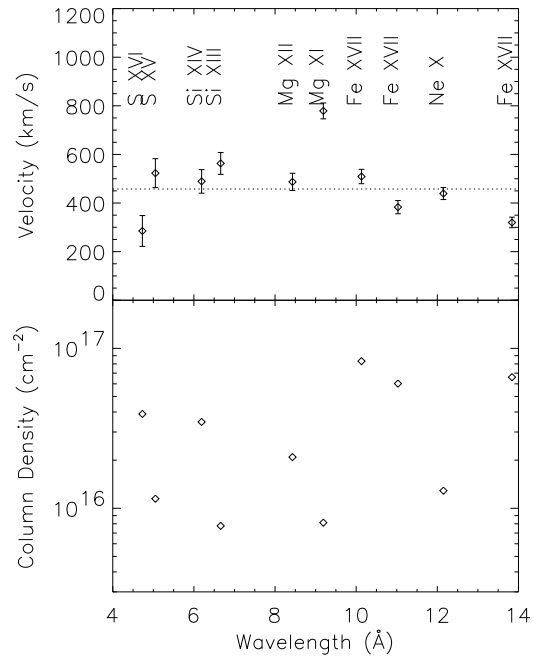,width=3.0in}}
\caption{The velocities and column densities of lines detected
in the Cyx X-1 X-ray spectrum as a function of their observed
wavelengths.  The dotted line in the top panel gives the
average redshift of 460 km/s, excluding Mg {\sc XI}.
}\label{fig:lines}
\end{figure}

Remarkably, we do not observe a wide spread of line
velocities (Fig.~\ref{fig:lines}), but 
a quite narrow ($ \pm 150$ km s$^{-1}$) distribution around 
450 km s$^{-1}$.
At the closest approach to the surface of the
companion, about one stellar radius, the line
of sight to Cyg X-1 will encounter a wind velocity of order
650 km s$^{-1}$.  In the focussed wind model, the velocity
vector is not directed strictly radially, so we observe
gas that appears to be on an accretion trajectory.
Using solar abundances, the column density of the absorbing
material is of order $10^{20-21}$ cm$^{-2}$, which
is well below the value of $10^{23}$ cm$^{-2}$ inferred
from \rxte\ ASM observations \cite{wen99}.
This could indicate that we observe either 
a physically narrow plasma flow (a ``streamer'' or ``ribbon'')
or only a smaller, more distant region in the plasma flow that
is either shadowed from the ionizing source or that is only 
partially ionized.

%do not change this

%do not change this
\normalsize

\section*{ACKNOWLEDGEMENTS}
 
This work has been supported in part under
NASA contracts NAS8-38249 and SAO SV1-61010.

\end{document}